\documentclass[onecolumn,referee]{aa}
\usepackage{txfonts}
\usepackage{graphicx}
\usepackage{natbib}
\bibpunct{(}{)}{;}{a}{}{,} %

\newcommand{\kms}{\rm ~km~s^{-1}}

\begin{document}

\title{Infrared Integral Field Spectroscopy of SN~1987A\footnote{Based on observations collected at the European Southern Observatory, Chile.}}
\titlerunning{IFU spectroscopy of SN~1987A}

\author{Karina Kj\ae r \inst{1}
\and Bruno Leibundgut \inst{1}
\and Claes Fransson \inst{2}
\and Per Gr\"oningsson \inst{2}
\and Jason Spyromilio \inst{1}
\and Markus Kissler-Patig\inst{1}}

\institute{ESO, Karl-Schwarzschild-Strasse 2, D--85748 Garching, Germany \\
\email{kekjaer@eso.org}
\and Dept. of Astronomy, Stockholm University, AlbaNova, SE-106 91
Stockholm, Sweden}

\date{Received / Accepted }

\abstract
{SN~1987A in the Large Magellanic Cloud is close enough for a study of the very late time evolution of a supernova and its transition to a supernova remnant. 
Nearly two decades after explosion we are witnessing the supernova shock wave engaging the inner circumstellar ring, which had been fluorescing since being ionised by the soft X-ray flash from shock breakout.}
{We follow the interaction of the supernova shock with the ring
material. The spatially resolved information provides us with insight
into the individual shock regions around the ring.  }
{Near-infrared integral field spectroscopy observations with
SINFONI/VLT of the SN-ring interaction is presented.
SINFONI's adaptive optics supported integral field spectrograph spatially resolves the ring and the data thus we obtain a better spatial understanding of the spectrum in different regions of the object.}
{With a dynamical map of the interacting ring we determine parameters
for its geometry. Since most of the IR emission lines originate behind
the shock front we obtain an indication of the radial velocity of the
shocked material after deconvolving the geometry. The ring geometry is
consistent with a circle and we also derive a new, independent
measurement of the systemic ring, and presumably also supernova,
velocity. We find from the
spatial distributions of the flux in the different emission lines the
degree of cooling in the shocked material and follows the increases
observed in the radio and X-rays. Emission from the ejecta is detected
only in the strongest [\ion{Fe}{ii}] lines.}
{}

\maketitle

\section{Introduction}

With the 8m telescopes in the Southern Hemisphere SN~1987A
will remain observable throughout its complete evolution from
supernova to supernova remnant. While the emission from the supernova
itself has been steadily fading \citep[e.g.][]{2003Bruno} the
inner circumstellar ring has remained bright. The collision of the
supernova shock with the circumstellar ring has been anticipated \citep{1994ApJLuo,1997ApJBork} and was first indicated by radio observations \citep{2001ApJBall,2002PASAManchester}. At about the same time the first hot spots
appeared just inside the circumstellar ring in 1996 showing that the supernova
shock had reached what are considered to be inward protrusions of the ring \citep[][]{1998ApJSonneborn,1998ApJMichael,2000ApJMichael,2002ApJMichael,2002ApJPun}. The X-ray emission started to increase as well \citep{2004ApJPark,2006ApJPark,2006AAHaberl}. The detection of coronal lines in the optical spectra from 2002 indicates that the ring material is shocked to a temperature of $\gtrsim 2 \times 10^6$ K \citep{2006AAPer}, consistent with the X-ray spectra \citep{2005ApJZhekov,2006ApJZhekov}.

The physics of the interaction is highly complex because of the nature
of the circumstellar medium of SN~1987A \citep{1995ApJChevalier,1999ApJLundqvist}. 

Here we present spatially resolved spectroscopy of the inner circumstellar ring from science verification observations with SINFONI on the VLT. 
Integral field spectroscopy yields both imaging and
spectroscopy, which allowed us to trace the shock interaction in
spatially separated emission sites around the ring. The approaching and
receding sides can be separated very clearly in several emission lines
and we are able to study the ring dynamics. 

The optical and IR provide, together with the UV and X-rays, the most
important diagnostics of the temperature, density, and ionisation of
the shocked gas. Therefore, the data presented here should be
seen as a complement to the high resolution optical and UV spectroscopy presented in \cite{2006AAPer,2007Per} and \cite{2002ApJPun}, as well as the X-rays by \cite{2005ApJZhekov,2006ApJZhekov}.

\section{Observations and Calibration}

We use SINFONI Science Verification
observations of SN~1987A carried out on 28 and 29 November
2004 \citep[day 6488, preliminary results were published in][]{2005MsngrGillessen}. Integral-field spectroscopy was obtained in J (1.11-1.35 $\mu$m), H (1.49-1.79 $\mu$m) and K~(1.99-2.38 $\mu$m) bands. 
The observations were AO supported using Star 3 as the reference source. We have not deconvolved the images, since deconvolving extended sources is unrealiable \citep{2007Davis}.
The pre-slit optics for SINFONI allows to chose different widths of the image slicing, which then leads to different fields of view on the sky. SN~1987A was observed with two different spatial resolution settings. 
The primary data in J, H, and K were obtained with a spatial resolution of 250 mas/spaxel and thus with a field of view of 8"$\times$8". In J we also obtained an observation with a spatial resolution of 100 mas/spaxel and a field of view of
3"$\times$3" (denoted as J100 in the following). All single
integrations were 600~seconds and Table~\ref{obsval} displays
the total integrations per setting. When several integrations
were obtained the individual exposures were offset by
sub-spaxel spacings to improve the spatial resolution. 
The exposures were separated by sky exposures of
equal integration time in an ABBA sequence.

\begin{figure}
\resizebox{\hsize}{!}{\includegraphics[width=4.5cm]{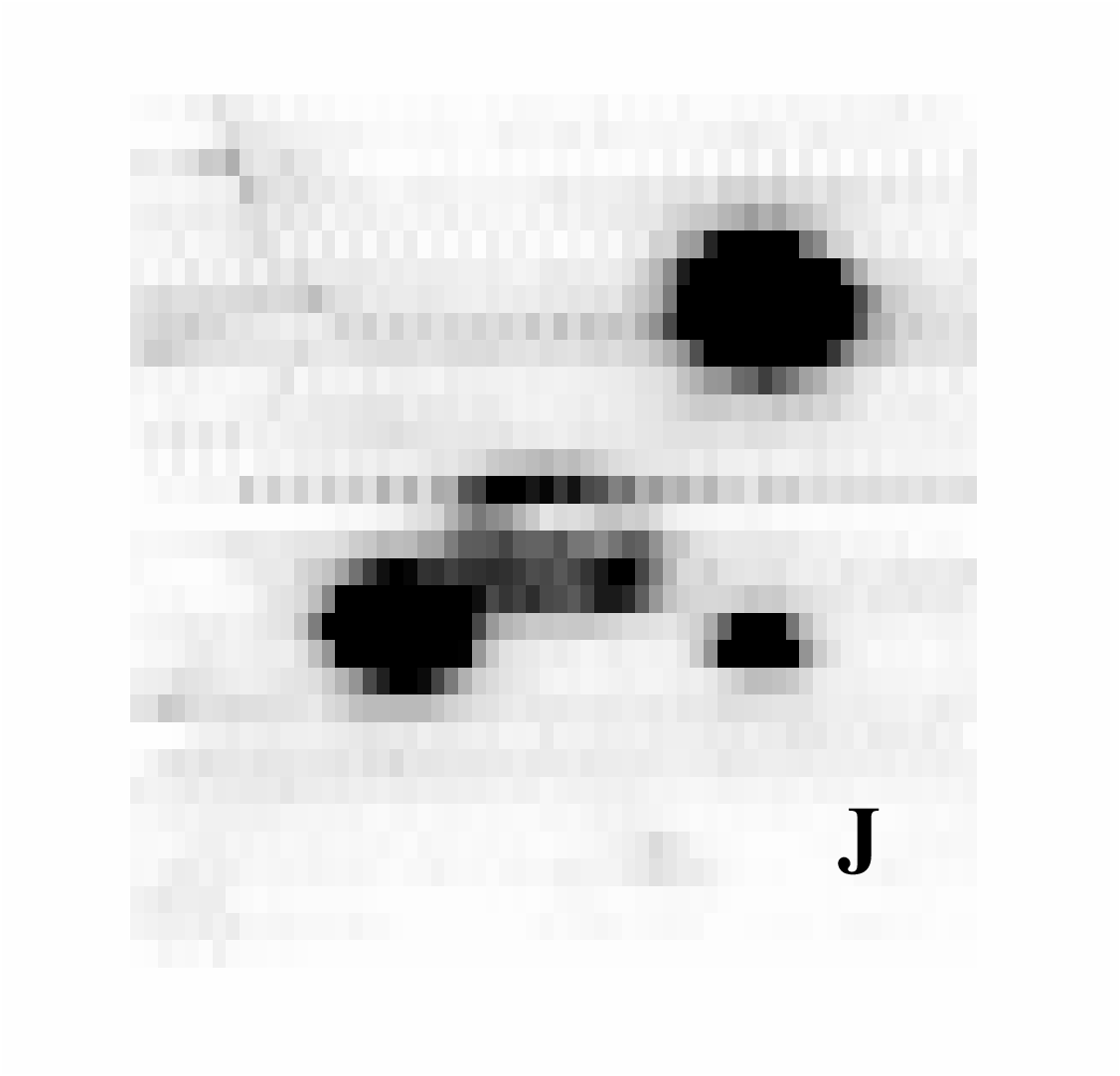}
\includegraphics[width=4.5cm]{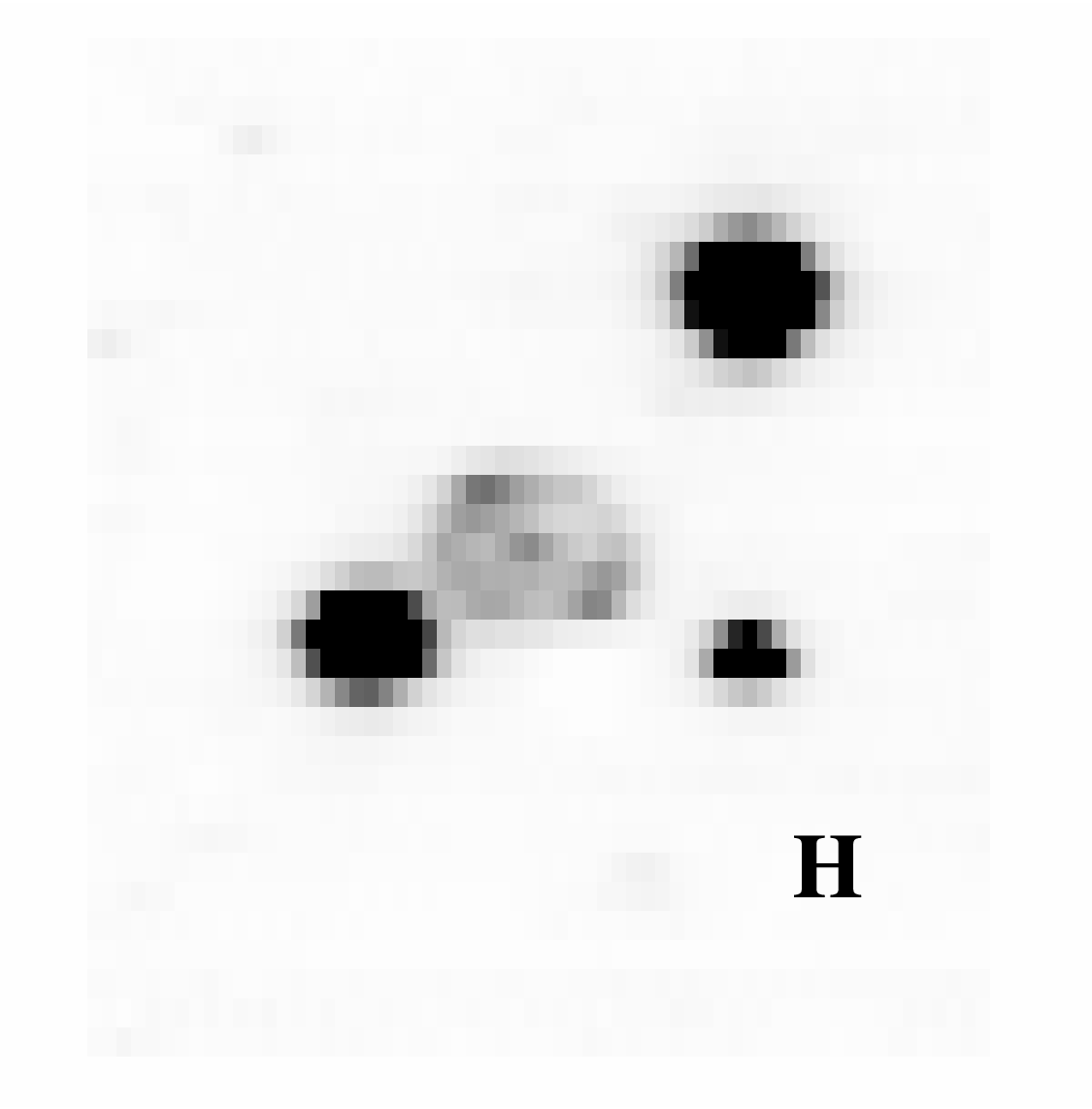}}
\resizebox{\hsize}{!}{\includegraphics[width=4.5cm]{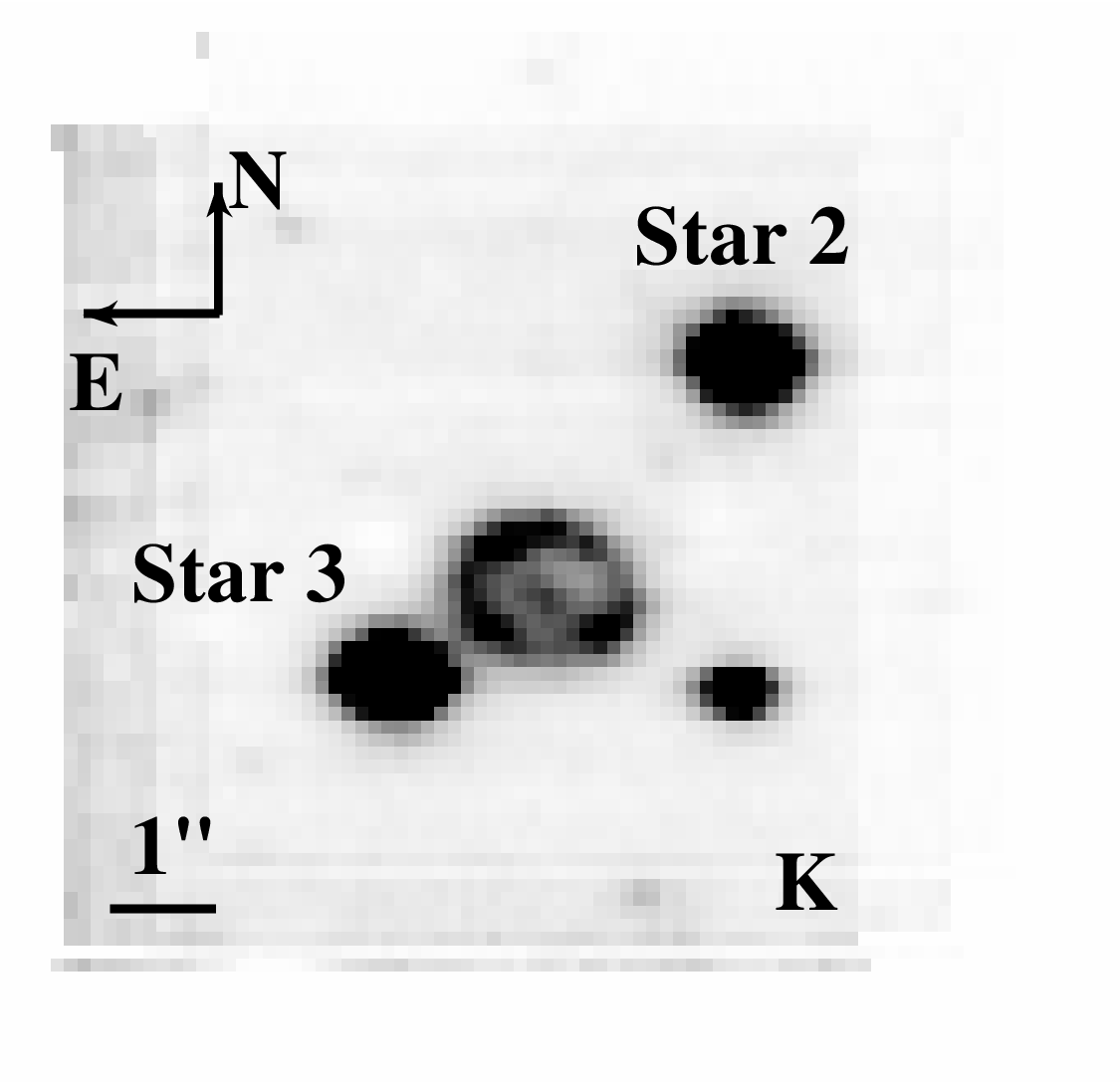}
\includegraphics[width=4.5cm]{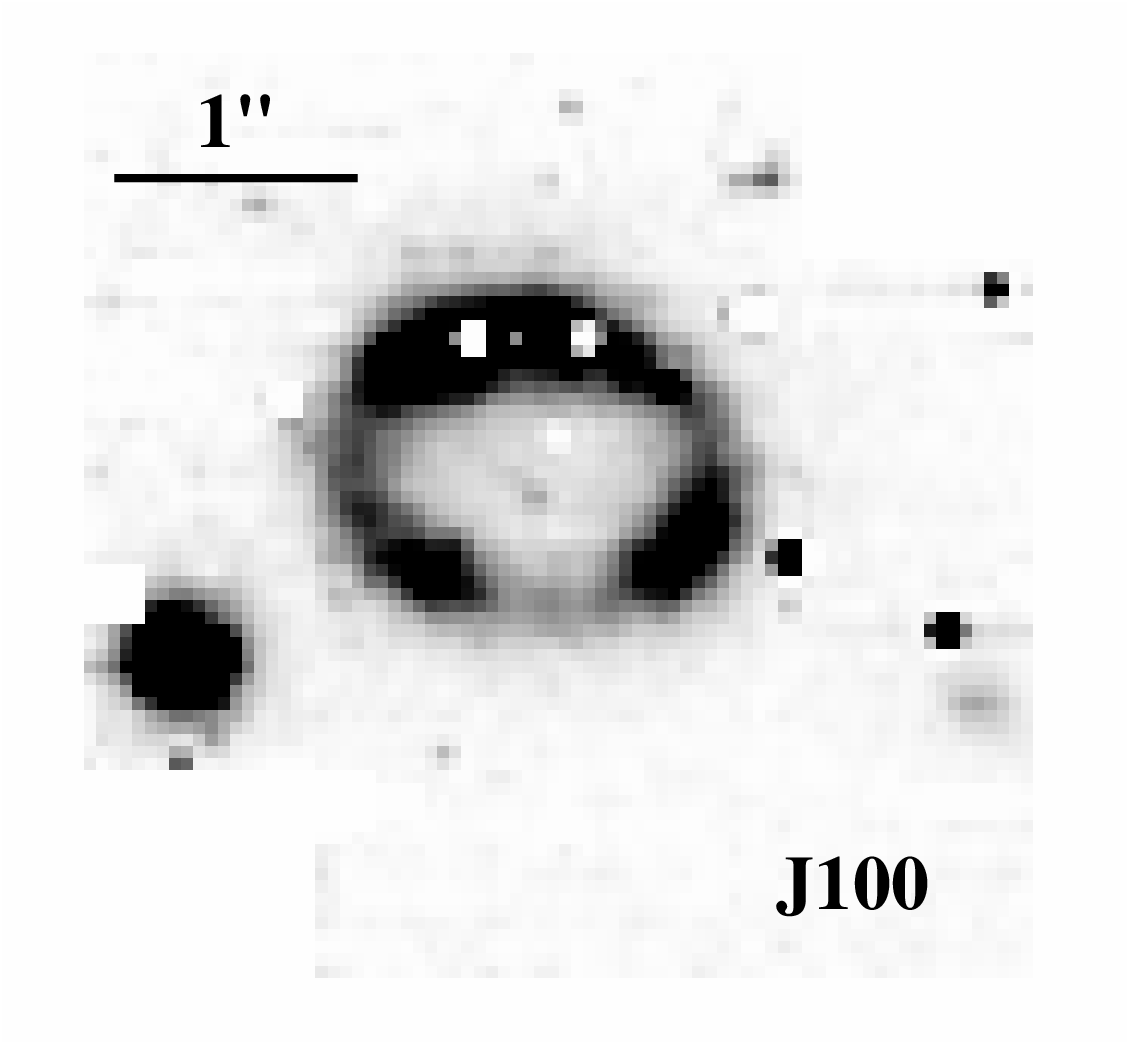}}
\caption{The 8''$\times$8'' FOV in the J, H and K-band images. Lower right is the higher resolution J-band 3''$\times$3'' FOV.}
\label{images}
\end{figure}

\begin{table} 
\caption{Summary of observations. The rows designate the observation
date (UT), the time since the explosion of SN~1987A on 23 February 1987A, the
spaxel size of the observation, the wavelength range of the
spectrograph setting, the spectral resolution, the total exposure time
and the image quality as measured from the combined data cube on Star 2 and Star 3. We
further list the standard star used for the flux calibration, and the
uncertainty in the flux as determined from a comparison of Star 2 from
the corresponding black body curve, the uncertainty in the wavelength
calibration as measured from night sky lines and the corresponding
uncertainty in velocity.}
\label{obsval}
\centering
\begin{tabular}{l|ccc|c}
\hline\hline
& J & H  & K & J100\\
\hline
Date & 2004/11/28& 2004/11/29& 2004/11/29& 2004/11/28\\
Day  & 6488 & 6489  & 6489 & 6488 \\
Spatial Res. & 250mas &250mas &250mas & 100mas\\
Wavelength & 1.11-1.35 $\mu$m & 1.487-1.792 $\mu$m& 1.985-2.375 $\mu$m& 1.11-1.35 $\mu$m \\
$\lambda/\Delta\lambda$&  2000 & 2000 & 4500 &  2000\\
Exptime & 600s & 1200s & 1800s & 1200s\\
Image quality & 0.62'' & 0.46'' & 0.48'' & 0.32'' \\
\hline
Std. Star&HD36531 &HD29303 &HD29303& HD36531\\
$\epsilon_{Flux}$ & 1 \% & 13 \%  & 4 \% & 4 \% \\
$\delta(\lambda)$ & 8 $\times 10^{-5}$ $\mu$m &  7 $\times 10^{-5}$ $\mu$m &  7.5 $\times 10^{-5}$ $\mu$m &8 $\times 10^{-5}$ $\mu$m \\
$\delta(v)$ & 19 km~s$^{-1}$& 12 km~s$^{-1}$ & 11 km~s$^{-1}$&19 km~s$^{-1}$ \\
\hline
\end{tabular}
\end{table}

The sizes of the spaxels for the observations are 250 mas
$\times$ 125 mas (and 100 mas $\times$ 50 mas for J100). In the K-band images the final image quality is superior to the J and H band due to better seeing and sampling. This can been seen in Figure \ref{images}, which shows the images (collapsed cubes) for the different bands. In the final combined data cube the spaxels are square and have the sizes 125 mas $\times$ 125 mas (50 mas $\times$ 50 mas for the J100). The ring is fully resolved in all bands. Using the stars in the FOV of the collapsed images we measure an image quality (values listed in Table~\ref{obsval}). The image quality measurement combines seeing, the AO correction and any smearing/dithering in combining the final cube. The image quality is in particular important in estimating reasonable sizes of spectral extractions (apertures).

The data have been reduced and
recombined into data cubes using the SINFONI pipeline \citep{2004ASPCSchreiber,2007pipeline}.  We crosschecked the wavelength solution
with the atmospheric OH lines \citep{2000AARousselot} and the resulting wavelength calibration is  accurate to 8$\times 10^{-5} \mu$m,
7$\times 10^{-5}\mu$m and 7.5$\times 10^{-5}\mu$m, in J, H and K,
respectively.  This corresponds to systematic velocity uncertainties
of 19~km~s$^{-1}$, 12~km~s$^{-1}$, and 11~km~s$^{-1}$, as listed in
Table~\ref{obsval}.

We dereddened the spectra using the galactic extinction law from \cite{1989ApJCardelli}, assuming $R_V=3.1$, and
$E_{B-V}$=0.16 \citep{1990AJFitzpatrick} for the color excess
towards SN~1987A, based on $E_{B-V}$=0.10 from the LMC and
$E_{B-V}$=0.06 from the Milky Way \citep{2003MNRASStaveley}. The
differences between the LMC and galactic extinction law are negligible
at low color excess in the near-IR. 

The standard stars were observed directly before or after the supernova and have been used to remove the telluric features. We used the
standard stars to construct a sensitivity curve and derive the flux
calibration, based on their measured magnitudes from the 2MASS catalogue \citep{2003.2MASS}. In this way the whole data cube had the same flux calibration,
and we could derive calibrated spectra for Stars~2 and 3 as well. To test the
absolute flux calibration we compared the spectrum of Star~2 \citep[a B0V star NW of the SN,][]{1993PASPWalborn} with the blackbody curve of a star with
R=8.3~R$_{\odot}$ and  T$_{eff}$=22500~K at the distance of 50~kpc. The derived
flux differences are of the order of 10\% and are listed as $\epsilon_{Flux}$ in Table~\ref{obsval}. For the J100 observations we calibrated the flux using Star 3, since Star 2 is not in this smaller FOV. As Star 3 is variable we used the observed values of Star 3 in the J250 to cross correlate the value. The total flux from Star 3 observed in J100 was 4\% different from the total flux from Star 3 obtained from J250 using the same aperture in arcseconds. We combined the two J observations into one spectrum weigthing the spectra with the exposure time. The flux error estimate for this combined spectrum is of the order of 2\%. We do not think that any emission contamination from the ring will influence the result beyond the 10\% we find as the accuracy of the flux.\\

\begin{figure}
\resizebox{\hsize}{!}{\includegraphics[height=8cm]{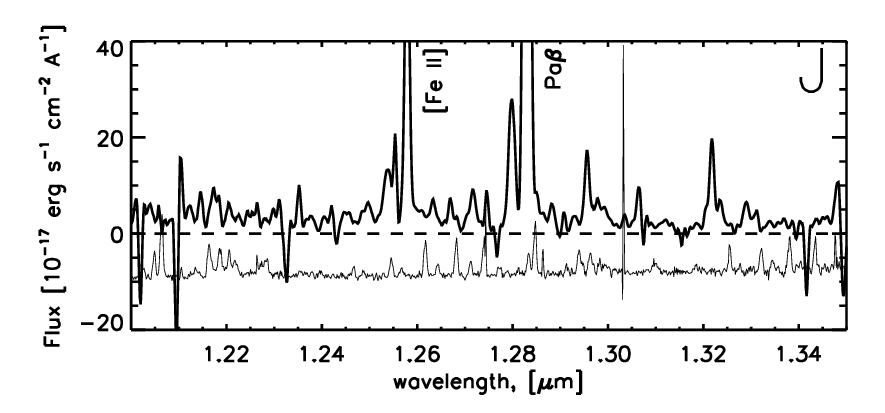}}
\resizebox{\hsize}{!}{\includegraphics[height=8cm]{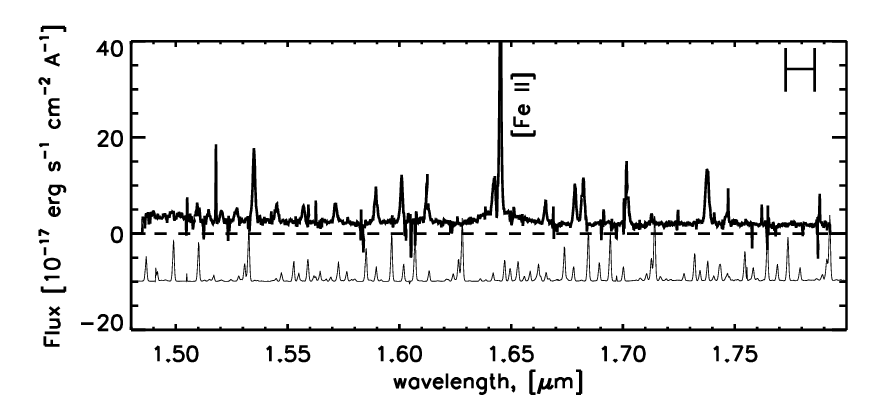}}
\resizebox{\hsize}{!}{\includegraphics[height=8cm]{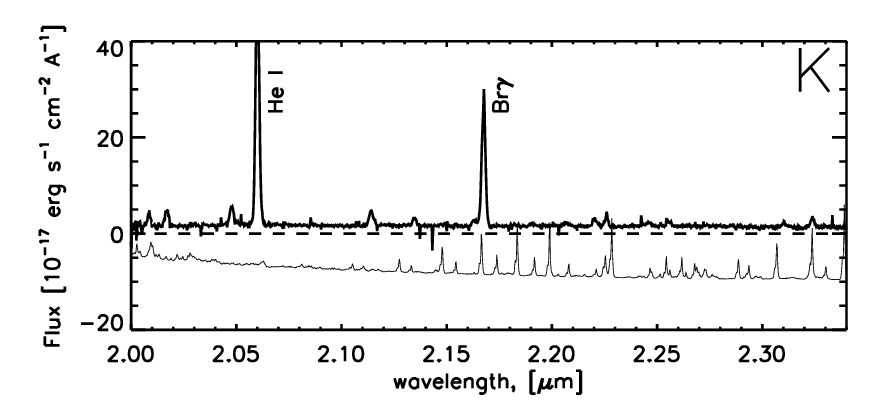}}
\caption{The integrated spectrum of the ring and the ejecta. The sky spectrum is shown offset below the SN ring spectrum. Note that the flux of the sky spectrum has been scaled down to make the spectrum fit in the same figure as the object. The line identifications are given in Table \ref{lineTJ}-\ref{lineTK}.}
\label{spectra}
\end{figure}

\section{Results}

\begin{table}
\caption{Emission lines of the integrated spectrum (ejecta and
circumstellar ring combined corresponding to the central 281 spaxels)
in the J~band. The spectral resolution is 150 km~s$^{-1}$ in the J band. Errors for the values listed in the table are given either in the text or in Table \ref{obsval}. }
\label{lineTJ}
\centering
\begin{tabular}{ccccccc}
\hline\hline
$\lambda_{obs}$& v$_{obs}$ & $\sigma_{obs}$ & $\sigma_{deconv}$ &  Flux & Identification & $\lambda_{air}$ \\
($\mu$m) & (km~s$^{-1}$) & (km~s$^{-1}$) & (km~s$^{-1}$) &
($10^{-16}$erg~s$^{-1}$~cm$^{-2}$) & & ($\mu$m) \\
1.253885 & 408 & 431& 404 & 14 & *[\ion{Fe}{ii}] a$^6$D$_{3/2}$-a$^4$D$_{1/2}$ & 1.2521\\
1.25385 & 261 & 431& 404 & 14 & *He I 1s3s-1s4p(3S-3P) & 1.2528\\
\multicolumn{7}{l}{Blend? The line transition probability for [\ion{Fe}{ii}] does not support this large flux. Furthermore large $\sigma_{deconv}$.}\\
\hline
1.25787 & 277 & 200& 132&69&$\dagger$[\ion{Fe}{ii}] a$^6$D$_{9/2}$-a$^4$D$_{7/2}$& 1.2567 \\
\hline
1.27169 & 315 & 236 & 182 & 7 & $\dagger$[\ion{Fe}{ii}] a$^6$D$_{1/2}$-a$^4$D$_{1/2}$& 1.2703 \\
\hline
1.27983 & 245 & 305 & 266 & 43 &$\dagger$[\ion{Fe}{ii}] a$^6$D$_{3/2}$-a$^4$D$_{3/2}$&1.2788\\
1.27983 & 310 & 305 & 266 & 43 & * He I 1s3d-1s5f(3D-3F) &  1.2785 \\
\multicolumn{7}{l}{Blend? The line transition probability for [\ion{Fe}{ii}] does not support this large flux. Furthermore large $\sigma_{deconv}$.}\\
\hline
1.28300 & 277  & 281 & 238 &  367 &$\dagger$Pa$\beta$  &1.2818   \\
\hline
1.29562 & 311  & 190 & 117 & 18  &$\dagger$[\ion{Fe}{ii}] a$^6$D$_{5/2}$-a$^4$D$_{5/2}$ &1.2943 \\
\hline
1.30651 & 302 & 253 & 204&12&  *\ion{N}{i}& 1.3052 \\
\hline
1.32178 & 277 & 166 & 71 & 23 & $\dagger$[\ion{Fe}{ii}] a$^6$D$_{7/2}$-a$^4$D$_{7/2}$ &1.3206 \\
\hline
1.32901 & 226 & 339 & 304 & 10 & $\dagger$[\ion{Fe}{ii}] a$^6$D$_{3/2}$-a$^4$D$_{5/2}$ & 1.3278\\
\hline
\multicolumn{7}{l}{The lines marked with $\dagger$ have been observed previously for this object in Meikle et al. (1993).}\\
\multicolumn{7}{l}{The lines marked with * are new for this object or have a new identification.}\\
\end{tabular}
\end{table}

Because the data format is 3-dimensional (spatial, spatial, and
wavelength) it is possible to integrate spectra over different
spatial regions. Generally, it is only correct to compare line
fluxes with respect to other line fluxes, where the number of co-added
spaxels, i.e. the aperture, is the same. In order to compare with previous data we summed
up the circumstellar ring and the ejecta (which corresponds to
281~spaxels co-added for the 250~mas data cubes). We extracted the
spectrum in a similar way for the J100 observations and scaled this
spectrum to the J spectrum before we combined the spaxels into a single
spectrum. 

We chose to scale the J100, because the aperture for the J
band is easily compared with the ones of the H and K bands, which were
observed with the same spatial scale. This spectrum should be comparable with earlier spectra of SN~1987A after a slit correction. We will refer to the spectrum
extracted this way as the 'integrated spectrum' and it is displayed in
Figure~\ref{spectra}. We also show a sky spectrum offset below the 'integrated spectrum'. This sky spectrum should be seen as a help to estimate where the more noisy regions are located in wavelength. The intensity of the various skylines cannot be directly compared with the integrated spectrum due to differences in aperture.

\begin{table}
\caption{Emission lines in the integrated spectrum (ejecta and
circumstellar ring combined corresponding to 281 spaxels) in H. The spectral resolution is 150 km~s$^{-1}$ in the H band. Errors for the values listed in the table are given either in the text or in Table \ref{obsval}.}
\label{lineTH}
\centering
\begin{tabular}{ccccccc}
\hline\hline
$\lambda_{obs}$& v$_{obs}$ & $\sigma_{obs}$ & $\sigma_{deconv}$ & Flux & Identification & $\lambda_{air}$ \\
$ (\mu$m) & (km~s$^{-1}$) & (km~s$^{-1}$) & (km~s$^{-1}$) & 
($10^{-16}$erg~s$^{-1}$~cm$^{-2}$) & & ($\mu$m) \\
\hline
1.49825 & 301  & 320 & 283 & 2   & *$\ion{H}{i}$ Br25& 1.4967\\
\hline
1.50986 &295  & 278 & 234 & 5  & $\dagger\ion{He}{i}$ 3s$^1$S-4p$^1$P$^0$ & 1.5084\\
\hline
1.51461 &  253  & 257 & 209 & 3 & * $\ion{H}{i}$ Br21  &   1.5133      \\
\hline
1.52037 & 232    & 276 & 232 & 3  & * $\ion{H}{i}$ Br20   & 1.5192   \\
\hline
1.52705 &  197   & 334 & 298  & 4  & *  $\ion{H}{i}$ Br19  &    1.5261   \\
\hline
1.53494 & 285 & 254 & 205 & 19& $\dagger[\ion{Fe}{ii}]$ a$^4$F$_{9/5}$-a$^4$D$_{5/2}$ & 1.5335\\
\hline
1.54512 & 237 & 272 & 227 & 5  & * $\ion{H}{i}$ Br17     & 1.5439 \\
\hline
1.55707 &273 & 270 & 224  & 5  &  * $\ion{H}{i}$ Br16 & 1.5556 \\
\hline
1.57142 &  257 & 305 & 266 & 7  & $\dagger\ion{H}{i}$ Br15 & 1.5701 \\
\hline
1.58947 & 273  & 283 & 240 & 10  &  $\dagger\ion{Si}{i}$ 4s$^1$P$^0$-4p$^1$P & 1.5880\\
& 267  & &&  &  $\dagger\ion{H}{i}$ Br14& 1.5881\\
\multicolumn{7}{l}{The transition probability for Br14 alone fits with the flux. \ion{Si}{i} identification tentative.}\\
\hline
1.60098 & 281  & 225 & 168 & 11   & $\dagger$[\ion{Fe}{ii}] a$^4$F$_{7/2}$-a$^4$D$_{3/2}$& 1.5995 \\
\hline
1.61232 & 257  & 279 & 235 & 12  & $\dagger\ion{H}{i}$ Br13 & 1.6109\\
\hline
1.64236 & 298  & 219 & 160 & 12   & $\dagger\ion{H}{i}$ Br12   &  1.6407 \\
\hline
1.64438 & 150 & 3393 & 3390 & 47   & $\dagger$[\ion{Fe}{ii}] a$^4$F$_{9/2}$-a$^4$D$_{7/2}$ &1.6435 \\
\multicolumn{7}{l}{Broad component. See text for comments. }\\
\hline
1.64518 & 296  & 171 & 82 & 51   &  $\dagger$[\ion{Fe}{ii}] a$^4$F$_{9/2}$-a$^4$D$_{7/2}$& 1.6435\\
\multicolumn{7}{l}{This component sits on top of the broad component (previous line).}\\
\hline
1.66534 & 282   & 234 & 180 & 6  &$\dagger$[\ion{Fe}{ii}] a$^4$F$_{5/2}$-a$^4$D$_{1/2}$& 1.6638\\
\hline
1.67851 & 290   & 233 & 178 & 12   &$\dagger$[\ion{Fe}{ii}] a$^4$F$_{7/2}$-a$^4$D$_{5/2}$ & 1.6769\\
\hline
1.68227 & 286   & 268 & 222 & 15   &  $\dagger\ion{H}{i}$ Br11  & 1.6807\\
\hline
1.70171 &  259  & 247 & 196 & 15   & $\dagger$\ion{He}{i} 3p$^3$P$^0$-4d$^3$D&1.7002\\
\hline
1.71289 & 307   &263 & 216  & 3  &$\dagger$[\ion{Fe}{ii}] a$^4$F$_{5/2}$-a$^4$D$_{3/2}$ & 1.7111\\
\hline
1.73769 & 254  & 311 & 272 & 21   & $\dagger\ion{H}{i}$ Br10& 1.7362\\
\hline
1.74661 & 286   & 327 & 291 & 6  & $\dagger$[\ion{Fe}{ii}] a$^4$F$_{3/2}$-a$^4$D$_{1/2}$& 1.7449\\
\hline
\multicolumn{7}{l}{The lines marked with $\dagger$ have been observed previously for this object.}\\
\multicolumn{7}{l}{The lines marked with * are new for this object or have a new identification.}\\
\end{tabular}
\end{table}

A very weak continuum flux can be discerned in the spectrum at $(2-3) \times
10^{-17} \mathrm{erg s^{-1} cm^{-2} \AA^{-1}}$. The consistency of this continuum
in all three bands makes us believe that it is real. The origin of this
continuum is not clear, but it could come from large scale emission in
the region. If it does indeed emerge from the ring system itself then it could be a weak H/He continuum. Most likely the continuum is a combination of several emission origins, and it is at the moment not possible to distinguish between them. 

\begin{table}
\caption{Emission lines in the complete spectrum (ejecta and
circumstellar ring combined corresponding to 281 spaxels) in K. The spectral resolution is 67 km~s$^{-1}$ in the K band. Errors for the values listed in the table are given either in the text or in Table \ref{obsval}.}
\label{lineTK}
\centering
\begin{tabular}{ccccccc}
\hline\hline
$\lambda_{obs}$& V$_{obs}$ &$\sigma_{obs}$ & $\sigma_{deconv}$ & Flux & Identification & $\lambda_{air}$ \\
($\mu$m) & (km~s$^{-1}$) & (km~s$^{-1}$) & (km~s$^{-1}$) &
($10^{-16}$erg~s$^{-1}$~cm$^{-2}$) & & ($\mu$m) \\
\hline
2.00855 & 275 & 239 & 229 & 5  &$\dagger$[\ion{Fe}{ii}] a$^4$P$_{1/2}$-a$^2$P$_{1/2}$ & 2.0067 \\
\hline
2.01696 & 272 & 327 & 317 & 8  & * [\ion{Fe}{ii}] a$^2$G$_{9/2}$-a$^2$H$_{1/2}$ & 2.0151 \\
\hline
2.04779 & 259 & 293 & 285 & 8  &$\dagger$[\ion{Fe}{ii}] a$^4$P$_{5/2}$-a$^2$P$_{3/2}$ &  2.0460 \\
\hline
2.06003 & 275 & 291 & 283 & 108 & $\dagger$\ion{He}{i} 2s$^1$S-2p$^1$P$^0$ & 2.0581\\
\hline
2.11412 & 279  & 369 & 363  & 8   & *\ion{He}{i} 1s3p-1s4s  & 2.1121 \\
\hline
2.13462 & 258 & 267 & 258 & 3  &  $\dagger$[\ion{Fe}{ii}] a$^4$P$_{3/2}$-a$^2$P$_{3/2}$ & 2.1328 \\
\hline
2.16331 &   & 416 & 411  & 4  & * &\\
\multicolumn{7}{l}{Which could be part of a broad component of Br$\gamma$ (not seen on the red side of Br$\gamma$).}\\
\hline
2.16754 & 277 & 291 & 283 & 58   &$\dagger\ion{H}{i}$ Br$\gamma$ & 2.1655 \\
\hline
2.19052 &  & 575 & 571 & 3   & * &\\
\hline
2.20659 & 255 & 381 & 375 & 2 & *\ion{Na}{i} $2p^65p- 2p^69s$ & 2.2047 \\
2.20659 & -58  & 381 & 375 & 2 & or $\dagger$ \ion{Na}{i} $4s^2S-4p^2P^o$ & 2.2070 \\
\hline
2.22026 &  & 365 & 359 & 4  & * &  \\
\hline
2.22593 & 290 & 243 & 234 & 5  & * [\ion{Fe}{ii}] a$^2$G$_{9/2}$-a$^2$H$_{11/2}$ & 2.2238 \\
\hline
2.24580 & 286 & 534 & 530 & 4  &  $\dagger$[\ion{Fe}{ii}] a$^4$P$_{1/2}$-a$^2$P$_{3/2}$&2.2436 \\
\multicolumn{7}{l}{Blend? The line transition probability for [\ion{Fe}{ii}] does not support this large flux. Furthermore large $\sigma_{deconv}$.}\\
\hline
2.25525 &  & 452 & 447 & 3  & * & \\
\hline
2.31030 & 304 & 403 & 397 & 3  & *\ion{Ni}{ii} $3p^63d^8(^3F)4s-3p^63d^8(^3F)4s$&  2.3079\\
\hline
2.32381 & 291 & 284 & 276 & 5  & *\ion{O}{i} $2s^22p^3(^2D°)3s-2s^22p^3(^4S°)5f$ & 2.3215 \\
\hline
\multicolumn{7}{l}{The lines marked with $\dagger$ have been observed previously for this object.}\\
\multicolumn{7}{l}{The lines marked with * are new for this object or have a new identification.}\\
\multicolumn{7}{l}{The lines 2.24-2.32 $\mu$m were observed in Meikle
et al. (1993) as a blended unidentified feature.}\\
\end{tabular}
\end{table}

The detected lines from the spectrum are listed in
Tables~\ref{lineTJ}, \ref{lineTH} and \ref{lineTK}. The weakest lines
we have been able to securely identify are those with a flux above 3$\times
10^{-16}$~erg~s$^{-1}$~cm$^{-2}$. In the tables $\lambda_{obs}$ denotes the
observed wavelength, $\lambda_{air}$ is rest wavelength in air for the
line transition listed under 'Identification'. The heliocentric velocity of a given line is listed in column v$_{obs}$ and calculated as the velocity shift between the observed wavelength and the rest wavelength for the identification and then corrected for heliocentric velocity. 
$\sigma_{obs}$ is the observed
FWHM of the line and $\sigma_{deconv}$ refers to the FWHM for the line
corrected for the instrument resolution ($\sigma_{deconv}=\sqrt{\sigma^2_{obs}-\sigma^2_{instr}}$). The integrated flux for the line is measured by
fitting a Gaussian to the line shapes and listed in the column labelled 'Flux' in the tables. We chose to ignore unresolved narrow lines, which we could identify as residue sky subtraction lines using the OH line catalogue of \cite{2000AARousselot}. Since we have a 3D data format we could investigate the spatial extent of each line in the integrated spectrum and check whether the line originated at a particular point in the object or whether it has a coherent appearance across the ring/ejecta.  This is a very effective way to check for artefacts, such as hot/bad pixels. All observed values are integrations over the whole ring.

We have used the NIST Atomic Spectra Database Version 3.1.0. \citep{NIST} to identify the lines, checking the following elements and their ions: H, He, C, N, O, Na, Mg, Si, S, Ca, Fe, Co, and Ni. The ions included in the NIST database are up to ionisation XXIV for the case of iron. The other elements are complete to a similar level. The few lines which we have not yet identified did not correspond to any of the mentioned elements using the NIST database version 3.1.0, which is not necessarily complete for these elements and all their ions. The accuracies of the wavelengths in NIST are implied by the number of significant figures. 

For the [\ion{Fe}{ii}] and \ion{H}{i} lines we have verified the identification by checking line ratios using the line transition probabilities from \cite{1988AANussbaumer} and the NIST database. The broad component in the H-band (Table \ref{lineTH}) is the [\ion{Fe}{ii}]1.644 $\mu$m line originating in the ejecta. The flux from the narrow component sitting on top of the broad component alone agrees well with the fluxes observed from the 1.257 $\mu$m and 1.321 $\mu$m [\ion{Fe}{ii}] lines, which come from the same upper level. However, if the broad component is the [\ion{Fe}{ii}]1.644 $\mu$m line then we should expect to see broad components from the lines [\ion{Fe}{ii}]1.257 $\mu$m and 1.321 $\mu$m as well. Given the S/N in our J observations they are not discernible. The two lines are located in a noisy region of the J-band especially when compared with the H-band region around the 1.644 $\mu$m line. This point can hopefully be resolved with higher signal to noise data.

When the identification transition in Tables \ref{lineTJ}-\ref{lineTK} is preceded with a dagger ($\dagger$), 
this transition has been observed previously and was identified by \cite{1993MNRASMeikle} or \cite{2002MNRAS.Fassia}.
Generally, the ring lines are now brighter than for those
observations, as expected if the shocked component dominates the narrow component from the unshocked ring. With the higher spectral resolution of SINFONI we can now
resolve some previously blended lines. However, there remain blends for lines with a true separation of less than 
$150 \kms$. We mark newly identified lines
with an asterisk (*) in the tables.

From HST spectroscopic observations \citep{2002ApJPun} and our
ground-based high-resolution spectroscopy \citep{2006AAPer,2007Per}, we know that the ring-ejecta system at the moment contains
three different velocity components. There is the narrow component (NC),
which displays a nearly Gaussian velocity distribution, with
$\sigma\sim 10~\kms$, arising from the unshocked circumstellar ring. This is
caused by fluorescence from the material recombining after being
ionised by the supernova's UV flash at shock breakout \citep{1989ApJFransson,1991ApJLundqvist}. The
intermediate component (IC), with $\sigma\sim 200-300 \kms$, originates
from the shocked ring material behind the passing shock front. The brightening has been observed in HST imaging \citep{1998ApJMichael,2000ApJMichael,2002ApJPun} and arises from this material, where the ejecta collides with the inner protrusions of the ring. In our observations the NC and IC cannot be distinguished and appear as a single marginally resolved component. The SN ejecta component with an expansion velocity of $\sim 15000 \kms$ comes from the interaction with the reverse shock \citep{2005ApJSmith,2006ApJHeng} not evident in our data. We also observe a broad component (BC) which will be discussed further in section \ref{broad}

\begin{figure}
\resizebox{\hsize}{!}{\includegraphics[height=8cm]{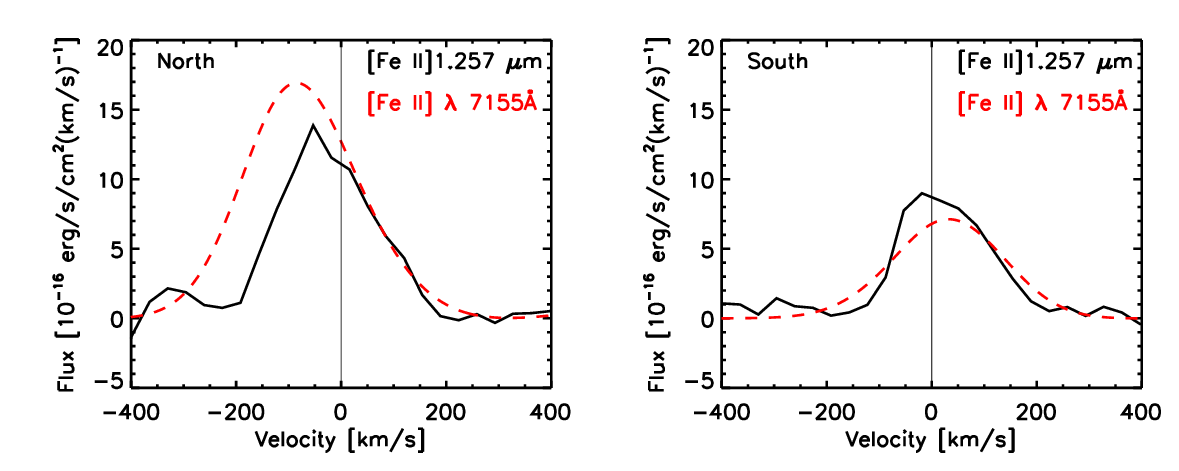}}
\caption{North and South extractions from SINFONI J100 of the [\ion{Fe}{ii}]1.257 $\mu$m, where the SINFONI data have been spatially smoothed to the image quality of the UVES observations, and extracted along a position angle (PA=30$^\circ$) of the UVES observations. The red/dashed curve are North/South extractions from UVES of the [\ion{Fe}{ii}]7155 $\AA$ line reduced to the spectral resolution of SINFONI scaled up a factor 100. } \label{uves}
\end{figure}

Because we do not have the spectral resolution to separate the narrow from the
intermediate-velocity component (cf. Tab.~\ref{obsval}) we expect some
of the intermediate lines we observed to be a convolution of the
narrow component with the intermediate component. In addition we have integrated spatially over the whole ring and thus the lines are also broadened by the velocity dispersion of the ring. Figure \ref{uves} shows a comparison of [\ion{Fe}{ii}] emission lines observed with UVES and SINFONI, where the UVES data has been scaled up by a factor of 100. The SINFONI observations have been convolved to the seeing of the UVES observations (0.5''). We extracted two spectra, one North and one South on the circumstellar ring emulating the 0.8'' slit (PA=30$^\circ$), which was the setup of the SN 1987A observations with UVES. The UVES spectra from the North and South part were convolved with the spectral resolution of SINFONI ($\lambda/\Delta\lambda \sim$2000 in the J-band) and shown in the figure as the dashed (red) curve. Taking into account that the wavelength accuracy in the J-band is $\pm$ 19 km/s then we have a good agreement for the peak velocity between the two observations. A detailed comparison of the individual elements with the high resolution optical spectrum \citep{2007Per} will further help in the separation of the different lines. 

\subsection{The Broad Component}
\label{broad}

\subsubsection{Central Emission}
\begin{figure}
\resizebox{\hsize}{!}{\includegraphics[height=8cm]{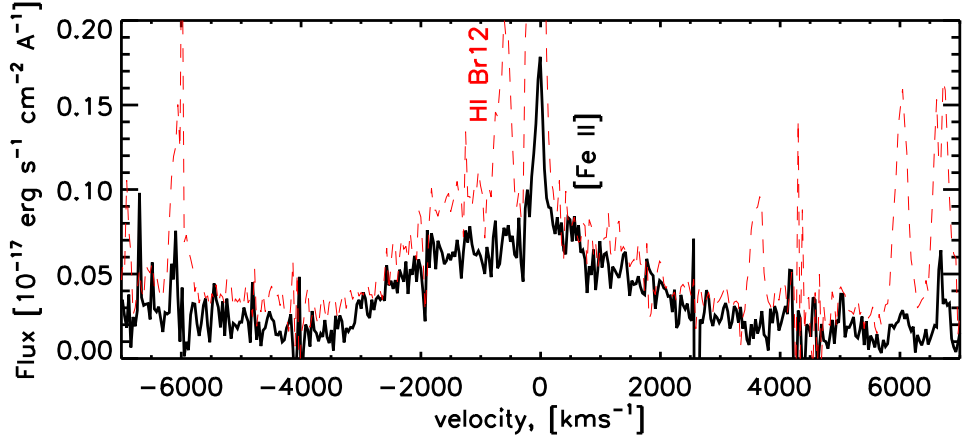}}
\caption{The [\ion{Fe}{ii}] $\lambda$1.644 $\mu$m line from a single spaxel centered on the ejecta at the center of the circumstellar ring (aperture=1 spaxel). The intermediate/narrow line on top is scatter originating in the ring. The red/dashed curve is the extraction at 25$^\circ$ (aperture area A=$\pi*(0.375'')^2$) scaled down with a factor of 10.}
\label{ejecta}
\end{figure}

The only line, for which we clearly can separate the broad ejecta component from the intermediate-velocity component in the integrated spectrum is the [\ion{Fe}{ii}] $\lambda$1.644 $\mu$m line. Figure~\ref{ejecta} shows this line from a single spaxel centered on the ejecta in the center of the circumstellar ring(full curve), for comparison we have plotted an extraction from Spot 1 (dashed line). The emission from Spot 1 has the intermediate Br12 line which is not present in the ejecta emission, furthermore there seems to be more blue emission in Spot 1 than in the ejecta.
 
For the integrated spectrum when we fit a Gaussian to both the broad part and the intermediate part of this line we find a small offset in the central wavelength, so that the broad line seems less red-shifted than the intermediate line, with respect to the rest frame of the system \citep[$v_{SN}=286.5$~kms$^{-1}$,][]{1995AAMeaburn}. The two components are individually listed in Table \ref{lineTH}. For the broad component we derived a velocity shift of 150~kms$^{-1}$ and a $\sigma_{deconv}\thickapprox$ 3400 km~s$^{-1}$. The bluer central wavelength for the ejecta component could indicate that the red part of the line is obscured by dust, as observed in the optical by \cite{1991Lucy}. 

This is also an indication on the faintness of the ejecta 17.8 years after the explosion. The ejecta spectrum is
now dominated by energy input from $^{44}$Ti \citep{2002NewFransson,2004ApJBouchet,2006ApJBouchet}. 

\subsubsection{Ring Emission}

\begin{figure}
\resizebox{\hsize}{!}
{\includegraphics[height=4cm]{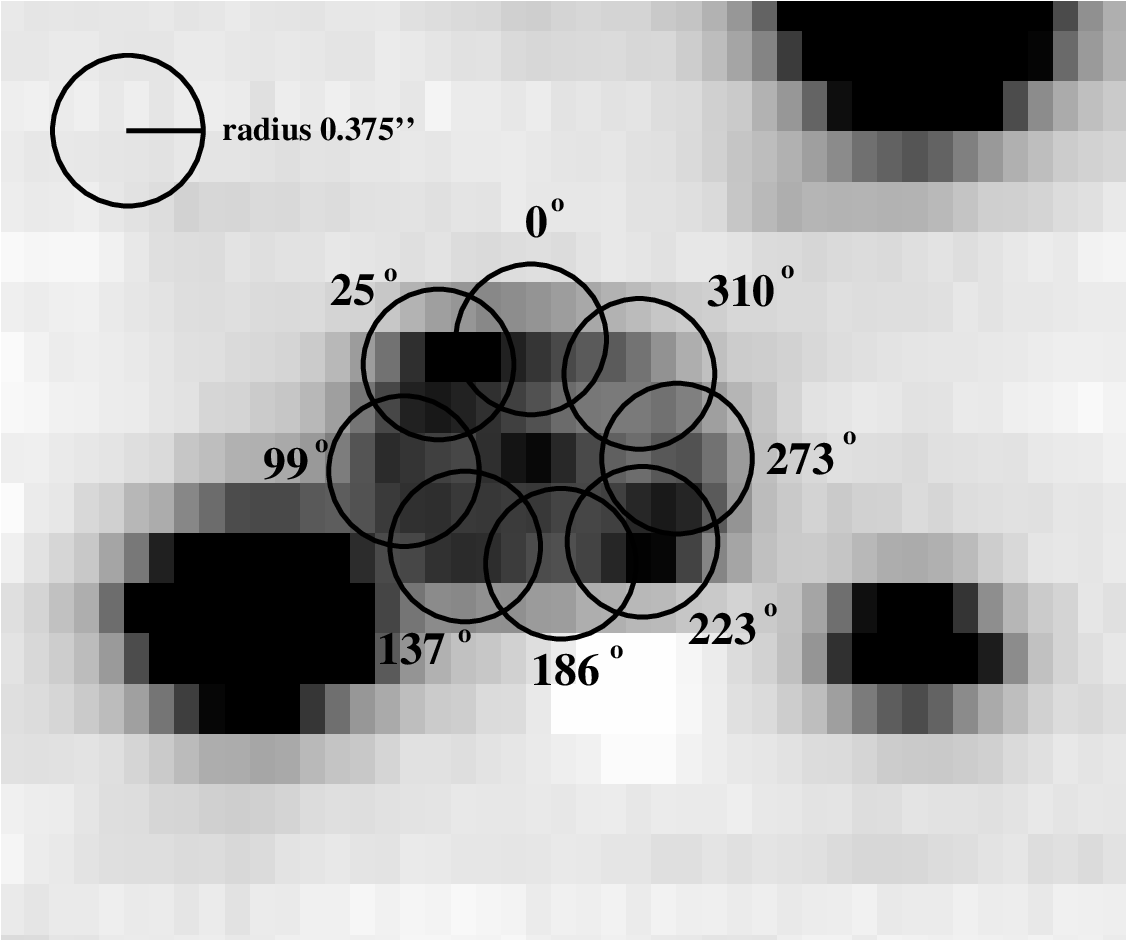}}
\caption{The size of the extractions and the corresponding azimuthal positions in the H band.}
\label{iron_extrations}
\end{figure}

\begin{figure}
\resizebox{\hsize}{!}
{\includegraphics[height=8.cm]{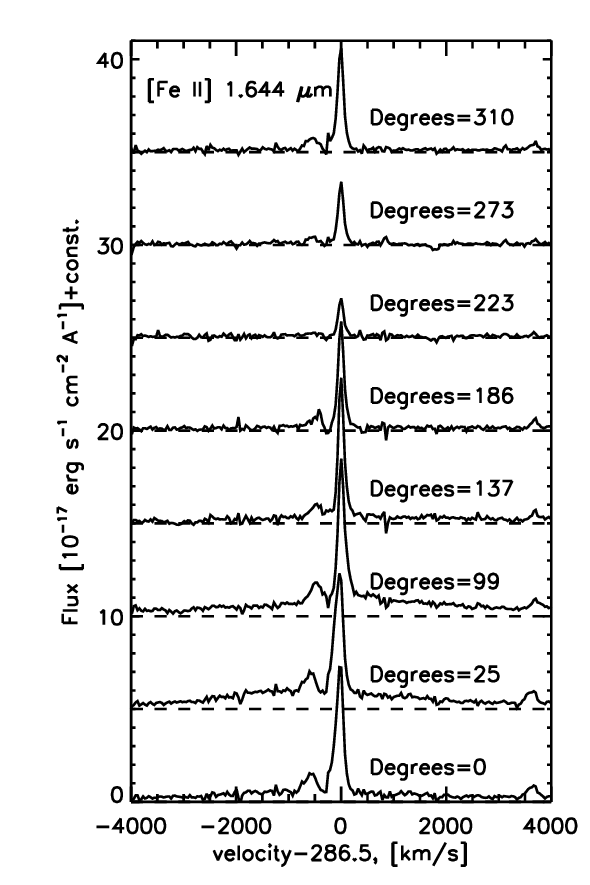}
\includegraphics[width=7.5cm]{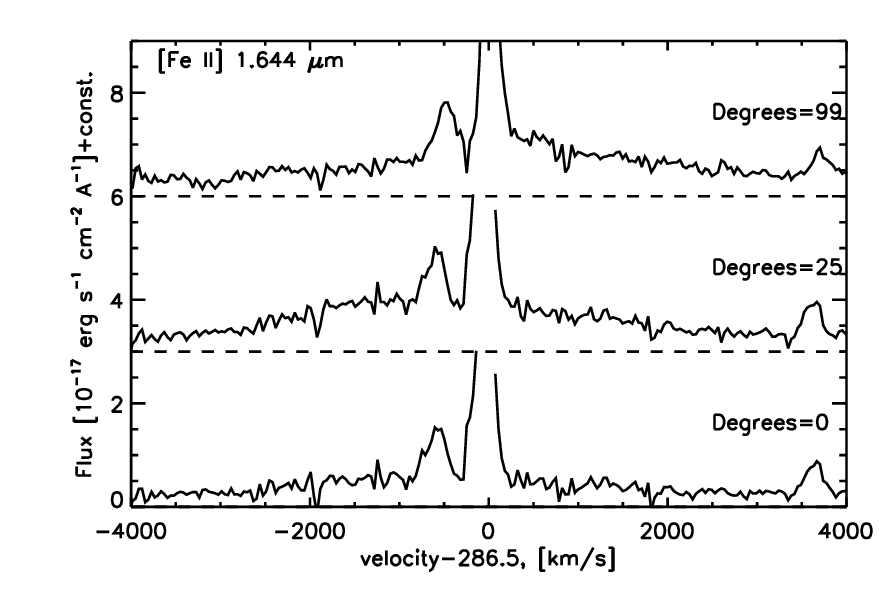}}
\caption{Extraction of spectra from the reduced data cube in the H band. The spectra are plotted with an offset. The faint emission on the blue side of [\ion{Fe}{ii}]1.644$\mu$m is Br12.}
\label{iron_spectra}
\end{figure}
We assigned an azimuthal angle for each position on the ring, defining North as zero degrees, and increasing the angle through East. Figure \ref{iron_extrations} shows the size of the extractions and their corresponding positions and azimuthal angles. Figure~\ref{iron_spectra} shows the extracted spectrum of [\ion{Fe}{ii}]1.644$\mu$m for different positions on the circumstellar ring and the zoom of three of the extractions. It is clear from this figure that the velocity shift of the [\ion{Fe}{ii}] 1.644 $\mu$m line's narrow component is not changing much for different azimuthal angles (different positions on the ring). Around Spot 1 and on the eastern side in general a broad component of the line emerges ($\sim$ 8000 km/s FWZI). The shape of the broad component (BC) changes even on the eastern side, with a slightly more luminous blue wing at 25 degrees, and a slightly more luminous red side at 99 degrees. However the ratio of blue to red flux in the broad component changes with azimuthal direction away from Spot 1 (see Fig. \ref{iron_spectra}). Therefore we conclude that the emission is not only scattered emission from Spot 1. Given the accuracy of this data set we cannot definitively exclude contamination.
Following the thorough discussion of the shocks in \cite{2002ApJPun} we think that the broad emission in the ring emerges from the reverse shock, which would be bluer in the North part of the ring due to the inclination of the ring with respect to the line of sight, which is consistent with our observations. Again, a deeper, high signal to noise observation would solve this question.

The broad line is only pronounced on the Eastern side of the ring. This East-West asymmetry has also been observed in X-rays and radio most recently in \cite{2006ApJPark} and \cite{2002PASAManchester}, respectively. The X-ray and radio emission is thought to emerge from the same region as the IR behind the shock front as the material cools down. We see also an East-West asymmetry with the intermediate lines, in that the Eastern side is brighter than the Western (see Fig. \ref{flux}).

Whether the East-West asymmetry is due to asymmetric ejecta outflow or differences in the density of the circumstellar matter (protrusions) is still an open question, which integral field spectroscopy observations of the reverse shock would shed light on. Since the hot spots primarily first appeared on the Eastern side we could also be witnessing spot evolution. This is a question which will be addressed in an upcoming paper which presents a higher spatial resolution data set for the epoch 2005.

\subsection{The Circumstellar Ring}
\label{resvel}
\begin{figure}
\resizebox{\hsize}{!}{\includegraphics[width=7.5cm]{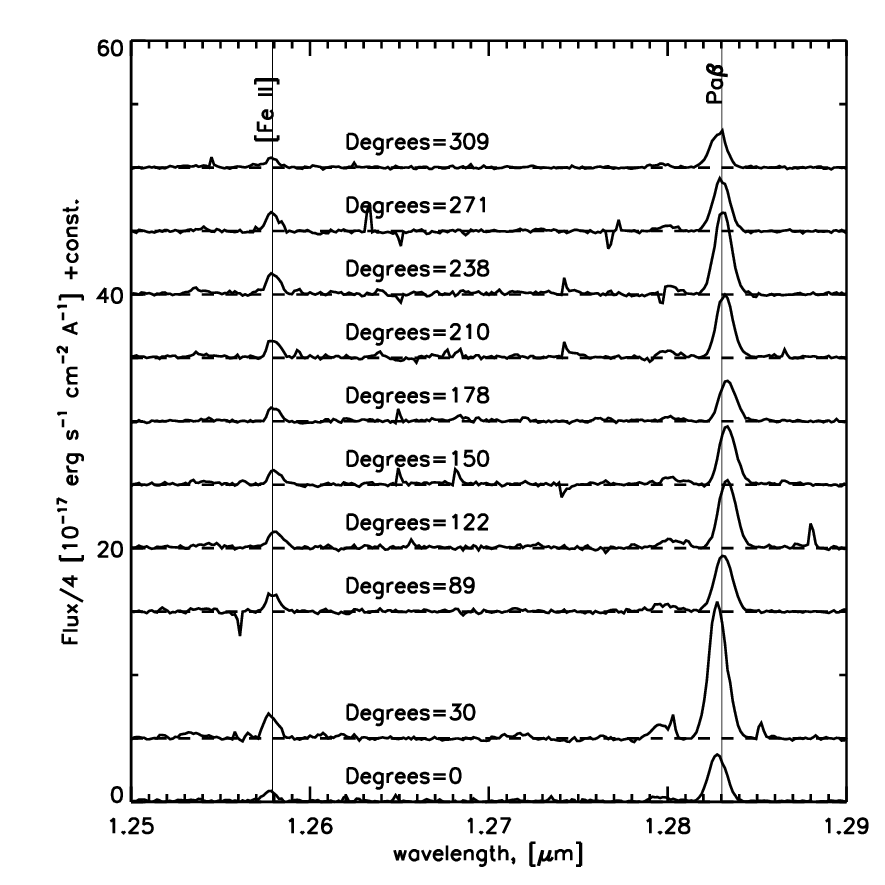}
\includegraphics[width=7.5cm]{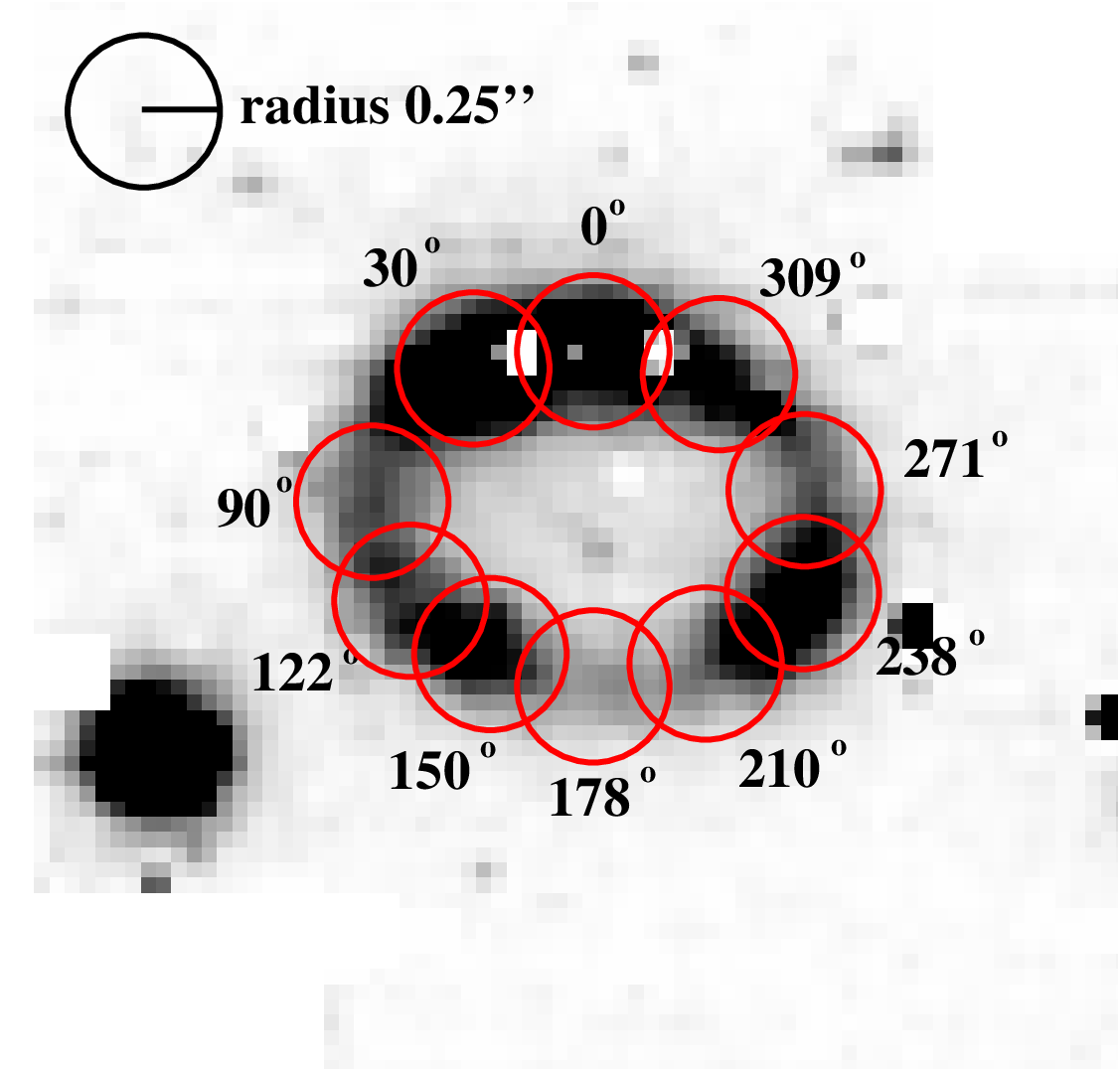}}
\caption{Left: Extraction of spectra from the reduced data cube close to the [\ion{Fe}{ii}] 1.257 $\mu$m and Pa$\beta$ lines in the J100 band. The spectra are plotted with an offset. Right: The size of the extractions and the corresponding azimutal positions in the J100 band. The vertical lines through the spectra mark the positions of the systemic velocity of 286.5~km/s for the different lines.}
\label{pab_extrations}
\end{figure}

The emission lines from the circumstellar ring are a complex
combination of different emission sites. With the shock moving into
the ring material an outward acceleration of the shocked material is
expected, and hence a difference in the projected radial velocity shift
around the ring should be observed. However, \cite{2000ApJMichael} and \cite{2002ApJPun} showed that there are oblique shocks along the protrutions and the velocity field is a combination of material accelerated not only radially forward. The SINFONI data cube can be used to explore the velocity structure of the ring. We assigned an
azimuthal angle for each position on the ring, defining North as zero
degrees, and increasing the angle through East. The resolution in
angle is then set by the spatial resolution. For each angle we coadded
a spectrum with the angular position as the center and the radius of
the order of the image quality. In this sampling two neighbouring extracted
spectra therefore have many spaxels in common and are correlated (cf. Fig. \ref{pab_extrations}). 
The radius of the extraction circle was 3 spaxels for the 250 mas
resolution (radius of 0.375'') 
and 5 spaxels for the 100 mas resolution (radius of 0.25''). We have here chosen the radius to be of the order of the image quality listed in Table \ref{obsval}, for the different observations.
The left panel of Figure \ref{pab_extrations} shows the extracted spectrum (J100-band between 1.25$\mu$m-1.29$\mu$m ) for different positions on the circumstellar ring and the right panel shows the sizes of the extractions and their corresponding positions and azimuthal angles. It is clear that the peak of [\ion{Fe}{ii}] 1.257 $\mu$m and the peak of Pa$\beta$ change for the different extractions.

\begin{figure}
\resizebox{\hsize}{!}{\includegraphics[height=8cm]{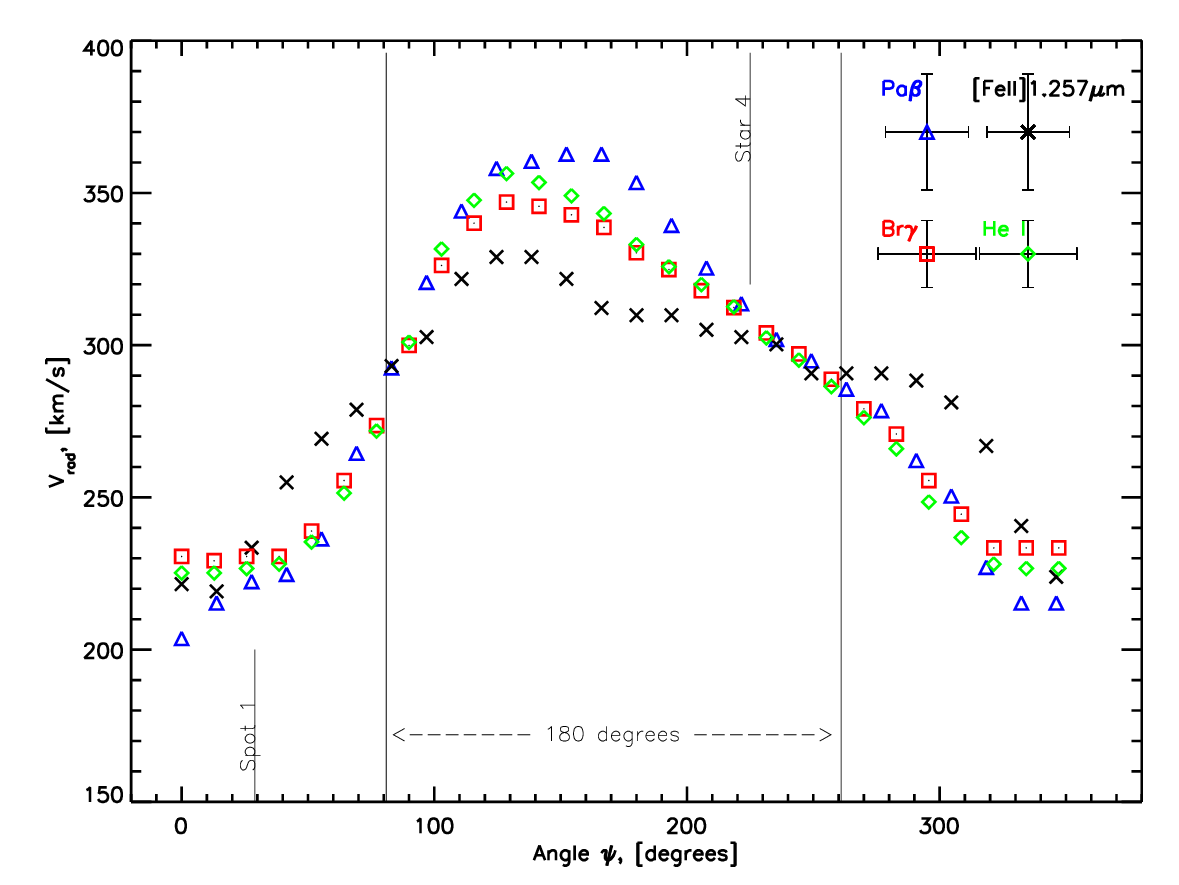}}
\caption{{Radial velocity shift for different angular positions around the ring for the strongest emission lines. The angle increases from North to East. The location of Spot~1 and Star~4 are indicated. The J100 data has been sampled with 7 px radius in order to have a similar extraction radius as the 250 obs. The errorbars are indicated in the top right corner. }}\label{vel}
\end{figure}

Figure~\ref{vel} shows the radial velocity shift for the Pa$\beta$, Br$\gamma$, \ion{He}{i}, and [\ion{Fe}{ii}]1.257$\mu$m lines, plotted $vs.$ the azimuthal
angle. We have here chosen the extraction radii for the J100 extractions to be similar to that of K250 (radius=7 spaxels $\sim$ 0.35''). The error bars in velocity are calculated from the uncertainty
in the determination of the wavelength and the error bars in angle display the size of the extraction
radius for the spectra. The error bars in angle are plotted conservatively to indicate the extent of contamination between adjacent spaxels. 

Only extractions with a distance of more than twice the integration radius are truly independent of each other. For J, H, and K the radius was 3 spaxels, which corresponds to a separation of 6 spaxels (0.75''), and 14 spaxels for J100 (0.7'').

The emission lines in Fig. ~\ref{vel} are all dominated by the
intermediate component, coming from the shocked ring material. This is
clear from the fact that the velocity of the lines differs by a large
factor from the expansion velocity of the narrow lines, $\sim 10 \kms$. It is
evident from this figure that the line centers of the northern part of
the ring are blue shifted, while the southern part is red shifted. This is as expected for an expanding circular ring with the Northern
part tilted towards us. From HST imaging \cite{1991ApJPanagia} find an
inclination angle i=$42.^\circ 8 \pm 2.^\circ 6$ with respect
to the plane of the sky \citep[see also][]{2005ApJSugerman}. The theoretical line of sight velocity,
$v_r$, for a circular ring, uniformly expanding with velocity $v_{exp}$, and tilted
with respect to the plane of the sky with the angle $i$, is

\begin{equation} 
v_r(\psi)= v_{exp} \sin i  \cos{(\psi+\phi)} + v_{SN} \label{vele} \ .
\end{equation} 
Here $\psi$ is the azimuthal angle of the ring segment, defined with
$\psi=0$ for North. The phase shift, $\phi$, is introduced to account for
the offset of the major axis of the ring from the East-West direction
and can be interpreted as a rotation of the ring out of the plane of the
sky. 

The observed velocity shift at a given angle is a complex convolution
of the individual shock velocities at this point and the geometry. A
simple model of the line profile of an individual spot was given in
\cite{2002ApJPun}. The key point of this model is that the shock
velocity is likely to vary along the surface of the hot spot depending on
the angle between the surface and the impacting blast wave. The
highest velocities are therefore expected for a head on collision,
while the tangential impacts result in lower shock
velocities. Furthermore, the gas velocity, in the reference system of the
shock, decreases behind the shock as the gas cools. For the observer
at rest, the hot gas velocity behind the shock is $3 V_s/4$, while
that of the radiatively cooled gas is $V_s$. The peak velocity at a
given angle therefore depends on the shock velocity which dominates the
contribution to a given line. It is therefore important to realise
that the peak velocity we measure is only a weighted
average of the emissivity of a given line. This is then modified by
the angle of the gas velocity to the line of sight.

This can clearly be seen in the VLT/UVES observations, which provide line
profiles with much higher spectral, but lower spatial resolution \citep{2006AAPer,2007Per}. The line profile of e.g.,
the Hydrogen Balmer and He I lines from the northern part of the ring, including Spot
1, has a peak at $\sim 200 \kms$ in November 2005. This agrees well
with that measured at the position of Spot 1 in our observations,
$\sim 223 \pm 7 \kms$ (combining Pa$\beta$, Br$\gamma$ and HeI 2.058 $\mu$m). The H$\alpha$ line in the UVES
observations, however, extends out to $\sim -450 \kms$, seen only as a
faint wing. While the UVES observations provide us with a good
representation of the emission as function of velocity, the SINFONI
observations measures the variation of the average velocity for each
line along the ring.  This illustrates well the complementarity of the
SINFONI and UVES observations.

It is clear from Fig.~\ref{vel} that the measured velocities do not
follow the simple cosine as predicted by the Eq.~(\ref{vele}). 

A simplistic way to derive the bulk velocity of the material, i.e. the
center of rest for the ring, would be to determine the 'nodal' points
($\psi=90^\circ$ and $\psi=270^\circ$), where the expansion velocity is orthogonal
to the line of sight. Inspection of Fig.~\ref{vel} immediately shows
that the two velocities are not identical at these angles for the various lines. 
We find that for Pa$\beta$, Br$\gamma$ and \ion{He}{i}
$v_r(90^\circ)=(300\pm7)$ km~s$^{-1}$ and $v_r(270^\circ)=(275\pm3)$
km~s$^{-1}$. For a perfect circular ring tilted out of the plane of the sky the
radial velocity should vanish at points exactly 180$^\circ$ apart and
these points indicate the bulk velocity of the material. We have
marked the two points where this happens in Fig.~\ref{vel}. The mean
bulk velocity is measured at a slightly different angle
$\psi+\phi=81^\circ \pm9^\circ$ and $\psi+\phi=261^\circ \pm 8^\circ$
with a bulk velocity of (280$\pm$7)~km~s$^{-1}$. This means that the
offset angle is $\phi =-9^\circ \pm 7^\circ$. This position of the major axis agrees well with that found by \cite{2005ApJSugerman} (PA= $81.^\circ1 \pm 0.^\circ8$) using imaging of light echoes and modelling the shape of the all the matter around the supernova. Our determination is not weigthed by intensity and rather than having single points we make a dynamical map of the whole ring. The influence of
the different hot spots or the contamination by Star~4, which is
directly superposed onto the inner stellar ring, appear negligible. We
have marked the location of Spot~1 and Star~4 in the diagram and there
are no clear deviations from the overall behaviour at these two
points. The derived velocity of the ring of (280$\pm$7)~km~s$^{-1}$ agrees with other results of 286.5 km/s \citep{2007Per,1995AAMeaburn,1995Crotts,1993Cumming}. 

\begin{figure}
\resizebox{\hsize}{!}{\includegraphics[height=8cm]{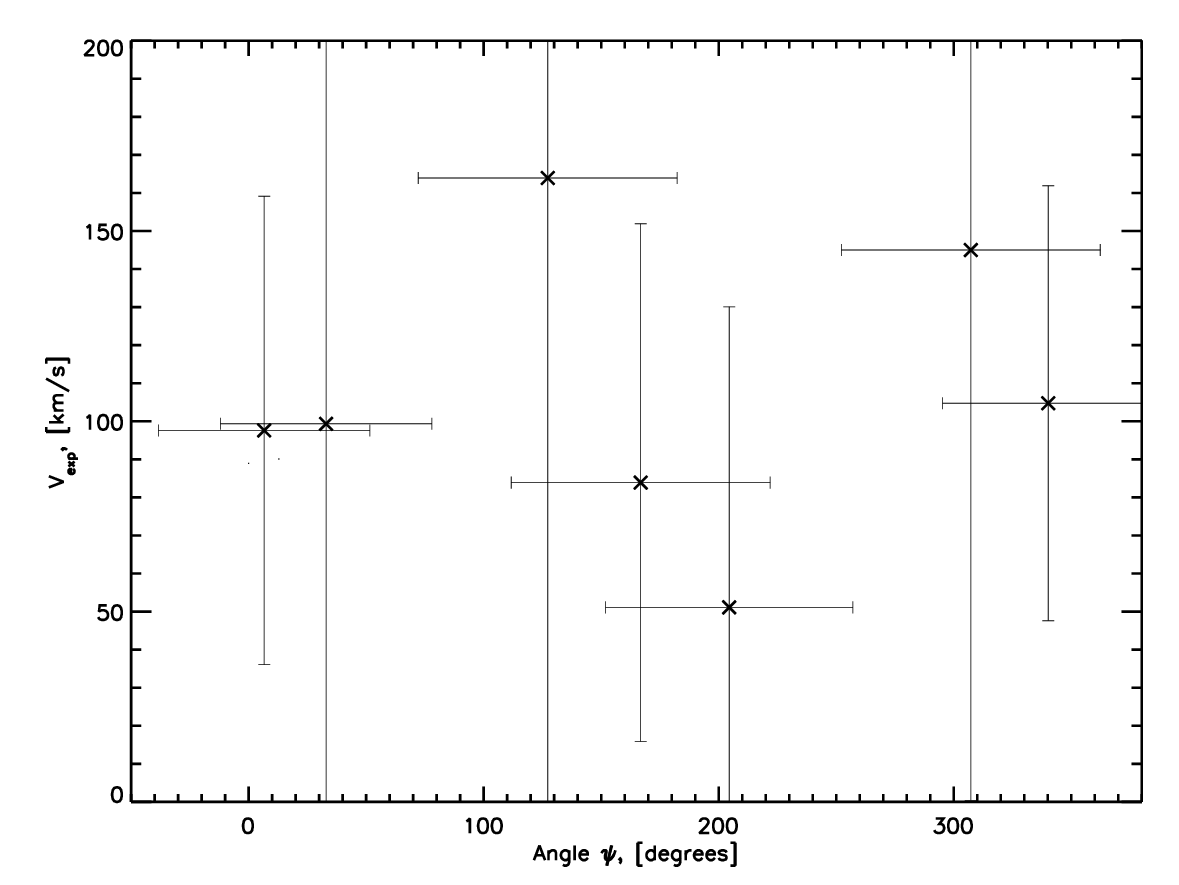}}
\caption{Deprojected velocity vs. angle for Pa$\beta$, Br$\gamma$ and
\ion{He}{i}. 0 degrees is North, the angle increases from North to East. The curves represents the propogated uncertainties for the deprojection, and we have cut the data where the uncertainty was larger
than the velocity measurement. In the deprojection we assumed a systemic velocity of 286.5 km/s.}
\label{exp_vel}
\end{figure}

Another result concerns velocity of the shocked ring material after
passage of the shock. As we discussed above, the peak velocity is a
function of the geometry of each shocked spot, as well as the
direction of the ring plane to the line of sight. We can, however, use
the geometric information above to find the deprojected, average shock
velocity around the ring. Assuming an inclination angle of
$i=-42^\circ.8$, and a systemic velocity of 286.5 $\kms$ we have calculated the expansion velocity and binned the datapoints in Fig.~\ref{exp_vel}. We derive from this a mean velocity around~$90 \kms$. 

\begin{figure}
\resizebox{\hsize}{!}{\includegraphics[height=8cm]{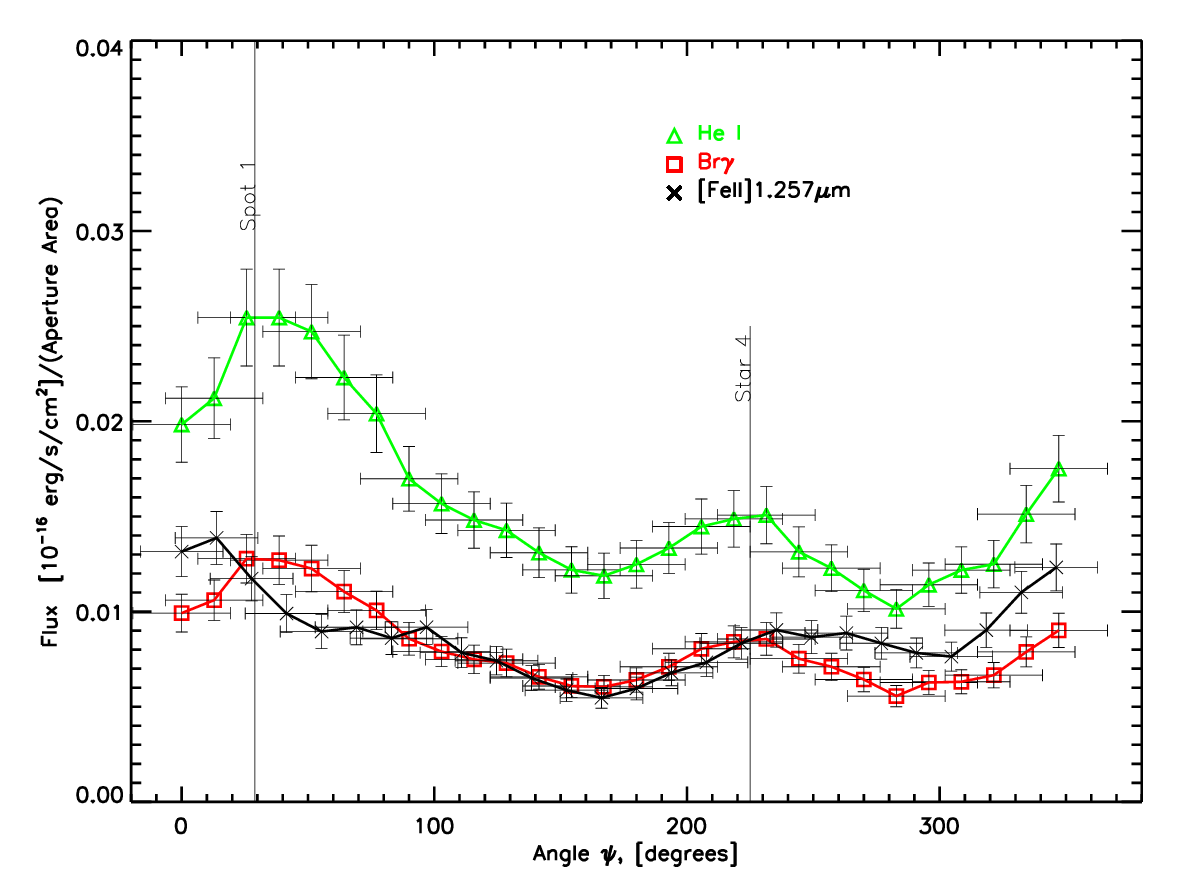}}
\caption{Flux variations for different angular positions around the
ring. The aperture area is A=$\pi*(0.375'')^2$ for the H and He lines, and A=$\pi*(0.35'')^2$ for the [\ion{Fe}{ii}] line.}
\label{flux}
\end{figure}

Figure \ref{flux} shows the flux of the [\ion{Fe}{ii}]
$\lambda$ 1.257 $\mu$m, the \ion{He}{i}, and the Br$\gamma$ lines
plotted vs. the azimuthal angle. We see here that the H and
He lines, as expected, follow each other, and that the ratio He/H is fairly uniform around the ring. While the narrow (intermediate) [\ion{Fe}{ii}] line flux follows the \ion{He}{i} flux level between 100$^\circ$ and 200$^\circ$ its flux increase around Spot 1 peaks at a different angle. This difference can be understood from the fact that while the H and He lines arise in an ionised region at $\sim 10,000$ K, the [Fe II] lines arise at $\sim 5000$ K, behind the ionized zone. The extent of this photoionized zone, and therefore the [Fe II] flux, depends on the shock velocity. In addition, collisional de-excitation is important
for the [Fe II] lines, while the H and He lines are less sensitive to
the density, being dominated by recombination. The ratio of the [Fe
II] and H and He lines will therefore depend on both the shock
velocity and density, which are both likely to vary along the
ring. Because $V_s \approx V_{\rm blast} (\rho_0/\rho_{\rm
blob})^{1/2}$ the shock velocity into the cloud depends on the cloud
density, $\rho_{\rm blob}$. Finally, if the ratio of the narrow and intermediate components vary between the different lines this may result in different peak velocities.

We expect the flux of H and He to increase around the ring as the hot spots evolve.

We will determine the evolution of the fluxes, as well as the velocity, around the ring in our next paper with epoch 2005 SINFONI data. 

\section{Summary}
 
The supernova shock interaction with the inner circumstellar ring
around SN~1987A is in full swing. Our integral-field observations have
revealed the extent of individual emission lines around the ring. From
the high-resolution optical spectroscopy of the coronal lines from UVES 
\citep{2006AAPer} and the soft X-ray component \citep{2005ApJZhekov,2006ApJZhekov}, we know that the shock is heating the ring material
to very high temperatures. The material behind the shock is cooling
rapidly, and the temporal
evolution of the lines will tell us more about the conditions in these
post shock areas. Our SINFONI Science Verification data only show
the strongest lines and we are now investigating deeper integrations
from a year later. The comparison of individual lines will be reserved
for a later paper.

Our data are still consistent with a circular ring, and we have derived
an average recession velocity, which agrees with other
measurements. Yet, we find that there are significant differences of
the velocities around the ring, with the eastern part showing larger
velocities than the western segment. This corresponds well with
the increase in surface brightness seen in both the radio observations
\citep{2002PASAManchester}, Chandra observations \citep{2006ApJPark} and the appearance of Spot 1 in the East \citep{2002ApJMichael,2002ApJPun,2002ApJSugerman}. By now the ring has been lit up in spots rather evenly
spaced around the ring, which we also observe in the emission lines of
H and He. 

From the velocity shift of the lines we find an average
velocity of $\sim 100 \kms$ for the bulk of the H and He emitting shocked gas.
We emphasise that this is just the average velocity, and considerably higher
velocity shocks are present, as is directly evidenced by the UVES
observations \citep{2006AAPer,2007Per}. 

The emission from the supernova itself has faded continuously. Today it is mostly the radioactive
decay of $^{44}$Ti and reverse shock interaction, which powers the emission from the ejecta. With
the integral-field spectrograph we can actually spatially isolate the
ejecta emission from the ring emission. The only clear line we detect
from the ejecta is [\ion{Fe}{ii}] 1.644~$\mu$m with a width of about
3400~km~s$^{-1}$. This is consistent with the radioactively and
reverse shock heated ejecta seen in H$\alpha$, Ca II and Mg I
\citep{2005ApJSmith,2007Per}. Deeper observations are likely to
detect more lines, as well as the broad wings from the reverse shock
in the H and He lines. An advantage of the IR lines is that the [\ion{Fe}{ii}] lines are less blended than in the optical spectrum. 

Monitoring the transformation of SN~1987A into a supernova remnant is
a unique opportunity. It is the only such object in which we can directly
observe this event while simultaneously resolving the interaction region. The structure of the circumstellar ring has been a puzzle since its discovery when it was
ionised by the SN soft X-ray flash. For a complete picture we need
to monitor this transition at all wavelengths. The SINFONI observations we are
obtaining are unique in that they allow us to spatially investigate
the emission sites around the ring and infer the local conditions of
the interaction.

\begin{acknowledgements}
We would like to thank the Garching and Paranal Astronomers
who provided support during the SINFONI Science Verification
runs. The observations could not have been carried out without
the help of Sabine Mengel, Thomas Szeifert and Christophe
Dumas.  We also would like to thank Andrea Modigliani for
valuable help in the data reduction. Special thanks go to
Maria Messineo, Mariya Lyubenova, and Nina Nowak for many
fruitful discussions regarding the science output from 3D
spectroscopy. This work was supported by the Swedish Research Council and the Swedish National Space Board (CF,PG).
\end{acknowledgements}

\bibliographystyle{aa}
\bibliography{SV_kjaer}

\end{document}